\documentclass[preprint,
superscriptaddress,nofootinbib,%
tightenlines ]{revtex4}
\usepackage{epsfig}
\usepackage{amssymb}
\usepackage{color}

\newcommand{\ben}{\begin{displaymath}}
\newcommand{\een}{\end{displaymath}}
\newcommand{\be}{\begin{equation}}
\newcommand{\ee}{\end{equation}}
\newcommand{\bea}{\begin{eqnarray}}
\newcommand{\eea}{\end{eqnarray}}
\begin{document}
\title{Critical comments on quantization of the angular momentum:\\
I.~Analysis based on the physical requirement on eigenfunctions and on the commutation relations}
\author{G.~Japaridze}
\affiliation
{Clark Atlanta University, Atlanta, GA, USA}
\affiliation
{Kennesaw State University, Kennesaw, GA, USA}
\author{A.~Khelashvili}
\affiliation{Institute of High Energy Physics, Iv. Javakhishvili Tbilisi State University, Tbilisi, Georgia}
\author{K.~Turashvili}
\affiliation{The University of Georgia, Tbilisi, Georgia}

\vspace{2in}


\begin{abstract}
Eigenfunctions and eigenvalues of the operator of the square of the angular momentum are studied. It is shown that neither from the requirement for the eigenfunctions be normalizable nor from the commutation relations it is possible to prove that the eigenvalues spectrum is a set of only integer numbers (in units $\hbar=1$). We present regular, normalizable eigenfunctions with the non-integer eigenvalues thus demonstrating that a non-integer angular momentum is admissible from the theoretical viewpoint. 

\end{abstract}




\maketitle

\section{introduction}

Quantization of the angular momentum is an important concept of the contemporary physics. 
In the framework of quantum mechanics derivation of quantization of the angular momentum is based on one of the following statements:
\begin{enumerate}
\item Quantization of the eigenvalues follows from the requirement that the eigenfunctions of the operator of angular momentum must be regular, i.e. normalizable \cite{shiff}.
\item Quantization follows from the commutation relations of the operators of physical quantities \cite{messiah}, \cite{wein}.
\item Quantization follows from the requirement that the eigenfunction of the third component of the operator of the angular momentum must be a single valued periodic function with the period $2\pi$ \cite{landau}, \cite{blokh}.
\end{enumerate}
All three derivations lead to the same result, namely that the spectrum of the angular momentum consists of only integer numbers (in units of Planck constant $\hbar$; throughout $\hbar=1$).


In this article we revisit derivation based on the first two statements and show that these derivations are based on mathematically not correct and not self consistent considerations. As a result, solutions with a non-integer spectrum become admissible on an equal footing as solutions with the integer spectrum.
Consequently, the statement that in the framework of quantum mechanics the 
eigenvalues of the square of the angular momentum and its third component are comprised of only integer numbers cannot be considered as strictly proven theoretical result.

Derivation based on the third statement will be analyzed in a subsequent publication where we will obtain the same result as in the present article, namely that the spectrum of the angular momentum may be comprised from both integers and non-integers.


For the clarity and comprehensibility of the arguments used in these two schemes of derivation it is convenient to give explicit analytic expressions of the eigenfunctions of the operators of angular momentum. For this reason we start with the discussion of details 
of solving the eigenvalue/eigenfunction equations for the angular momentum. 

The article is organized as follows: In section \ref{secII} we discuss properties of eigenfunctions of the operator of the square of angular momentum. In section \ref{secIII} we analyze mathematical arguments based on which, when solving for the eigenfunctions of the square of the angular momentum, 
it is argued that the spectrum consists only of integers. We point out inaccuracy  in using these arguments. In section \ref{secIV} we discuss the commutation relations of 
angular momentum operators from which the spectrum of integer eigenvalues of the square of the angular momentum is obtained. Using the results of section \ref{secII} we indicate the 
mathematical fallacy which leads to only the integer spectrum.

Main result of this article is that the solution of the eigenvalue problem of the orbital angular momentum contains physically admissible regular, i.e. normalizable eigenfunctions with the eigenvalues of the operator of angular momentum integer as well as non-integer.

Our conclusions are summarised in section \ref{secVI}.

\section{Regular and singular eigenfunctions of the operator of the square of the angular momentum}
\label{secII}
The eigenvalue equation for the square of the angular momentum takes its simplest form in spherical coordinates and reads:
\begin{eqnarray}
\hat M^2 \psi (\theta,\phi) &=& \left[ {1\over \sin\theta}  \,\frac{\partial}{\partial\theta} \left( \sin\theta \,\frac{\partial}{\partial\theta}\right)
+ {1\over \sin^{2}\theta}\,\frac{\partial^2}{\partial\phi^2} +\lambda \right]
\psi(\theta,\phi)=0,
\label{eq1}
\end{eqnarray}
where $\theta,\;\phi$ are spherical coordinates, $0\leq \theta \leq \pi, \ \ 0\leq \phi < 2\pi$, $\hat M^2 $ is the operator of the square of the angular momentum, $\lambda$ is its eigenvalue which we write as $\lambda=L(L+1)$, and without loss of generality we assume that $L\geq 0$.
Due to the commutativity of the operators of the square of the angular momentum and its third component $M_z=-i\partial/\partial \phi$
solutions to Eq.~(\ref{eq1}) can be written as a product $\psi(\theta;\phi)=\Psi_{Mm}(\theta;\phi |L;m)=\Psi_M(\xi | L;m)\Psi_m(\phi)$ the factors of which satisfy following equations:
\begin{eqnarray}
&& \hat M_z \Psi_m (\phi) = m \Psi_m (\phi) , \\
&& (1-\xi^2){d^2 \Psi_M\over d \xi^2}-2 \xi \,{d \Psi_M\over d \xi} -\left( \frac{m^2}{1-\xi^2} - \lambda\right )\,\Psi_M=0, \label{eq1.4}
\end{eqnarray}
where $\xi\equiv\cos\theta,\;-1\leq\xi\leq 1$ and $m$ is the value of the third component of the angular momentum.

The set of eigenfunctions  $\Psi_M = \Psi_M(\xi | L;m)$ consists of subsets  of regular and singular functions, regular being those that 
have no singularities within the domain of $\xi$. Our aim is to identify these subsets.

It is convenient to present solution in the form $\Psi_M(\xi)=(1-\xi^2)^\beta F(\xi)$. Substituting in Eq.~(\ref{eq1.4}), setting $\beta^2=m^2/4$ and $\xi^2\equiv z$ we bring Eq.~(\ref{eq1.4})
to the standard form of the Gauss hypergeometric equation 
(see, e.g., Eq.~15.5.1 in Ref.~\cite{Abramowitz})
\begin{eqnarray}
&& z (1- z) {d^2 F\over dz^2} + [c-z( a+b+1)] {dF\over dz}- a b F \nonumber\\
&& \ \ \ = z (1- z) {d^2F\over dz^2}+ [1/2-z( 3/2+2 \beta)] {dF\over dz}+[\lambda/4-\beta/2-\beta^2] F = 0 ,\label{gia}
\end{eqnarray}
where 
\begin{eqnarray}
&& \ \ \ a =[1/2+2\beta+(1/4+\lambda)^{1/2}]/2,\ \ b =[1/2+2\beta - (1/4+\lambda)^{1/2}]/2; \ \ c = 1/2.
\label{eq6}
\end{eqnarray}
This equation has two linearly independent solutions (see Eqs.~15.5.3-4 of Ref.~\cite{Abramowitz}):
\begin{eqnarray}
&& {}_2 F_1 (a,b;1/2;z) = {}_2F_1(1/2+\beta+L/2,\beta-L/2;1/2;\xi^2), \\
&& z^{1/2}\, {}_2 F_1 (a+1/2,b+1/2;3/2;z) =\xi \; {}_2F_1(1+\beta+L/2,1/2+\beta-L/2;3/2;\xi^2) ,
\label{eq712}
\end{eqnarray}
where ${}_2F_1(a,b;c;\xi^2)$ is the Gaus's hypergeometric function \cite{Abramowitz}.
For the three possible values, $2\beta =\sqrt{m^2}=\{ |m|; +m,-m\}$, three different expressions are obtained for $\Psi_M(\xi | L;\beta)$. On the other hand, as the original equation (\ref{eq1.4}) depends on $m$ 
only quadratically, all three parametrisation of $\beta$ must lead to the same result. To demonstrate this invariance, let us give explicit expressions for $\Psi_M(\xi|L;\beta)$, the two linearly independent solutions of Eq.~(\ref{eq1.4}):
\begin{eqnarray}
\Psi^0_M(L;\beta)&=& (1-\xi^2)^{\beta} { }_2 F_1 \left( 1/2+\beta+L/2, \beta-L/2; \frac{1}{2};\xi^2\right), \label{first}\\
\Psi^1_M(L;\beta)&=& \xi (1-\xi^2)^{\beta} { }_2 F_1 \left( 1+\beta+L/2, 1/2+\beta-L/2; \frac{3}{2};\xi^2\right).
\label{eq8}
\end{eqnarray}
$\Psi^0_M(\xi)$ is an even function of $\xi$ and $\Psi^1_M(\xi)$ is an odd function of $\xi$. As mentioned above both functions must be invariant under the change of the sign of $m$. For $2\beta=|m|$ the invariance is explicit. 
For $2\beta=\{ m;-m\}$ 
the invariance is not obvious but it can verified by using the following relation (see Eq.~15.3.3 of Ref.~\cite{Abramowitz}):
\begin{equation}
{}_2F_1(a,b;c;z)=(1-z)^{c-a-b}{}_2F_1(c-a,c-b;c;z)
\nonumber
\end{equation}
Using this relation it is straightforward to show that both $\Psi^0_M$ and $\Psi^1_M$ are even functions of $\beta$:
\begin{eqnarray}
\Psi^0_M(L;\beta)=\Psi^0_M(L; - \beta), \quad \Psi^1_M(L;\beta)= \Psi^1_M(L; -\beta).
\label{eq8}
\end{eqnarray}
Thus if some result is obtained in any one parameterization, then the same result can be obtained also in any other parameterisations. These parameterisations lead to different 
degrees of complication in calculations, therefore we should use the most convenient form for the representation of the corresponding functions. 

To single out the subset of normalisable functions let us study the singularities of functions $\Psi_M(L;\beta)$. These functions can be singular only for $\xi^2=1$. 
For example, for $\beta\geq 0$ the factor $(1-\xi^2)^\beta$ in $\Psi_M(L;\beta)$ is regular and the hypegeometric functions may have singularities of the order $(1-\xi^2)^{-\beta-\varepsilon}$, $\varepsilon > 0$, leading to the singular solution $\sim (1-\xi^2)^{-\varepsilon}$.
However, if the parameters of the hypergeometric function ${}_2F_1(a,b;c;z)$ satisfy conditions $a=-k$ or $b=-k$, where $k$ is a non-negative integer, then this hypergeometric function turns into the $k$-th order polynomial 
of $z$ \cite{Abramowitz}. Correspondingly, in this case hypergeometric functions will have no singularities. This conditions of truncating hypergeometric series, i.e. reducing hypergeometric functions into polynomials can be used to single out the subset of normalisable functions from the set of the solutions of Eq.~(\ref{eq1.4}).

As an example let us identify the regular functions for the solutions $\Psi_M(L;\beta)=\Psi^0_M(L;m/2)$, parameterization $2\beta=m$. 
We have two independent conditions for terminating infinite hypergeometric series (\ref{first}), thus reducing it to polynomials:
\begin{eqnarray}
&& a=\frac{1}{2}+\frac{m}{2}+\frac{L}{2}=-k\;\rightarrow \; m=-L-1-2k,\nonumber\\
&& \Psi^0_M(L;m)|_{m=-L-1-2k}= (1-\xi^2)^{-\frac{(L+1)}{2}-k} \ { }_2 F_1 \left( - \frac{1}{2}-L-k, -k; \frac{1}{2};\xi^2\right), \label{firstt}
\end{eqnarray}
and
\begin{eqnarray}
&& b=\frac{m}{2}-\frac{L}{2} =-k\;\rightarrow\; m=L-2k,\nonumber \\ 
&& \Psi^0_M(L;m)|_{m=L-2k}= (1-\xi^2)^{\frac{L}{2}-k} \ { }_2 F_1 \left( -k,\frac{1}{2}+L-k; \frac{1}{2};\xi^2\right).
\label{eq10}
\end{eqnarray}
Function obtained from the first condition (\ref{firstt}) is singular for any non-negative integer $k$ because the exponent of $(1-\xi^2)^{-(L+1)/2-k}$ is a negative number and the second factor, the hypergeometric function is a polynomial and hence is a regular function of $\xi$. 
The second condition (\ref{eq10}) leads to singular as well as regular subsets of functions. In particular, for $L/2-k\geq 0$, i.e. for $k<[L/2]$, where $[L/2]$ is an integer part of $L/2$ (remember that $k$ is integer), under condition that $0\leq L/2-[L/2]< 1$, both factors in Eq.~(\ref{eq10}) are regular. 
For $L/2-k<0$ the factor $(1-\xi^2)^{L/2-k}$ is singular and hence $\Psi^0_M(L;m)|_{m=L-2k}$ is also singular. 
We obtain that the eigenfunction, corresponding to the spectrum $m=-L-1-2k$, is singular:
\begin{equation} 
\Psi^0_M(L;m)|_{m=-L-1-2k}= \Psi^{0,S}_M(L;-L-1-2k),
\label{eq11}
\end{equation}
and the set of eigenvalues $m=L-2k$ factorizes in two subsets:
\begin{eqnarray}
&& m=L-2k= m|^L_{(L-[L])} {\rm U} \, m|_{-\infty} ^{(L-[L]-2)},\nonumber\\
&& \Psi^0_M(L;m)|_{m=L-2k\geq 0}= \Psi^{0,R}_M\left(L;m|^L_{(L-[L])} \right),\nonumber\\
&& \Psi^0_M(L;m)|_{m=L-2k < 0}= \Psi^{0,S}_M(L;m|_{-\infty} ^{(L-[L]-2)}).
\label{eq12}
\end{eqnarray}
Here $A|^{A_{max}}_{A_{min}}\,{\rm U}\,B|^{B_{max}}_{B_{min}}$ stands for the union of sets $A$ and $B$, and in notations of functions the index $S$ indicates singularity of the corresponding functions and index $R$ - the regular character of the corresponding functions. 

For the parameterization $2\beta=-m$ conditions for $\Psi^0_M$ being regular mirror those of Eq.~(\ref{eq12}). Instead of (\ref{firstt}) and (\ref{eq10}) we now have:
\begin{eqnarray}
&& a=\frac{1}{2}-\frac{m}{2}+\frac{L}{2}=-k \;\rightarrow\; m=L+1+2k,\nonumber\\
&& \Psi^0_M(L; -m)|_{m=L+1+ 2k} = (1-\xi^2)^{-\frac{(L+1)}{2}-k} \ { }_2 F_1 \left( - \frac{1}{2}-L-k, -k; \frac{1}{2};\xi^2\right), \label{firsttt}
\end{eqnarray}
and
\begin{eqnarray}
&& b=-\frac{m}{2}-\frac{L}{2} =-k\;\rightarrow\; m=-L+2k,\nonumber \\ 
&& \Psi^0_M(L;-m)|_{m=-L+2k}= (1-\xi^2)^{\frac{L}{2}-k} \ { }_2 F_1 \left( -k,\frac{1}{2}+L-k; \frac{1}{2};\xi^2\right),
\label{eq13}
\end{eqnarray}
That is, although the functions coincide with those of Eqs. (\ref{firstt}), (\ref{eq10}) respectively, the spectrum of $m$, determined by the condition of getting polynomials, mirrors the spectrum (\ref{eq12}):
\begin{eqnarray}
&& m=L+1+2k; \ \ \Psi^0_M(L;m)|_{m=L+1+2k}= \Psi^{0,S}_M(L;L+1+2k),\nonumber\\
&& m=-L+2k= m|_{-L}^{(-L+[L])} \, {\rm U} \, m|^{\infty}_{(-L+[L]+2)},\nonumber\\
&& \Psi^0_M(L;-m)|_{m=-L+2k\leq 0}= \Psi^{0,R}_M\left(L;-m|_{-L}^{(-L+[L])} \right),\nonumber\\
&& \Psi^0_M(L;-m)|_{m=-L+2k > 0}= \Psi^{0,S}_M\left(L;-m |^{\infty}_{(-L+[L]+2)}\right).
\label{eq1415}
\end{eqnarray}
We conclude that in the set of eigenfunctions $\Psi^0_M(L;\pm m)$ the subset of regular functions is given by the following spectrum of $m$:
\begin{eqnarray}
&& m^{(R)}= m|_{-L}^{(-L+[L])} \, {\rm U} \, m|^{L}_{(L-[L])},\nonumber\\
&& m|_{-L}^{(-L+[L])} = \{ -L;-L+2; \cdots;-L+[L]\}; \ \ m|^{L}_{(L-[L])} = \{ L;L-2; \cdots;L-[L]\},\nonumber\\
&& m^{(R)}= \{ -L;-L+2; \cdots;-L+[L]; L-[L]; \cdots; L-2;L\}.
\label{eq16}
\end{eqnarray}
The rest of the spectrum of $m$ which consists of values 
\begin{eqnarray}
&& m^{(S1)} = \{ -\infty;\cdots; -L-3;-L-1\}; \ \ m^{(S2)} = \{ -\infty; \cdots; L-[L]-4;L-2[L/2]-2 \},\nonumber\\
&& m^{(S3)}= \{ L+1;L+3; \cdots;\infty\}; \ \ m^{(S4)}= \{ -L+[L]+2;-L+[L]+4; \cdots; \infty\},
\label{eq17}
\end{eqnarray}
corresponds to the subset of singular eigenfunctions in the set of eigenfunctions $\Psi^0_M(L;\pm m)$. 

Same procedure is used for the second linearly independent function $\Psi^1_M$, for which we just
state the result. Regular functions and the corresponding spectrum have the form:
\begin{eqnarray}
&& m^{(R)}= m|_{-L+1}^{(-L+1+[L-1])} \, {\rm U} \, m|^{L-1}_{(L-1-[L-1])} \nonumber\\
&& = \{ -L+1;-L+3; \cdots;-L+1+[L-1]; L-1-[L-1]; \cdots;L-3; L-1\},\nonumber\\
&& \Psi^{1,R}_M\left(L;m^{(R)}\right)= \Psi^{1,R}_M\left(L; -m^{(R)}\right)=\xi(1-\xi^2)^{\frac{L-1}{2}-k}{}_2F_1\left( -k, \frac{1}{2}+L+k; \frac{3}{2}; \xi^2\right).
\label{eq18}
\end{eqnarray}
The subset of singular eigenfunctions and the corresponding spectrum are obtained in same way as for $\Psi^0_M$.

Finally, the subset of regular eigenfunctions and the corresponding spectrum can be described as follows:
\begin{enumerate}
\item
In the set of the two linearly independent solutions to the eigenvalue problem of the square of the angular momentum the subsets of regular eigenfunctions are generated by the mutually independent conditions of reducing hypergeometric functions to polynomials of $\cos\theta$.

\item

The subset $\Psi^{0,R}_M(L;m^{(R)})$ of linearly independent regular eigenfunctions corresponds to the spectrum of $m$ 
which is symmetric under the reflection of the sign: $m^{(R)}=m_0^{(R)}=m_0^{(-R)} \, {\rm U} \, m_0^{(+R)}$;

\begin{enumerate}
\item
The subsets are labeled by spectrum of $m$ which is a numeric sequence with the step size 2, $|m_j-m_{j-1}|=2$, (in units of $\hbar$), satisfying condition $m_0^{(-R)}=-m_0^{(+R)}$.

\item
The minimal value in the set $m_0^{(R)}$ is $m_{0min}^{(R)}=m_{0min}^{(-R)}=-L$ and the maximal value is $m_{0max}^{(R)}=m_{0max}^{(+R)}=L$.

\item
If $L$ is an integer, moving through spectra of $m$ with the above mentioned step size 2 we transit from  subset labeled by $m_0^{(-R)}$ to subset labeled by $m_0^{(+R)}$ and vice versa. In other words, these subsets are continuations of each other.


If $L$ is not an integer, moving through spectra of $m$ with the above mentioned step size 2 does not lead to the transition from one subset to another, i.e. in this case the subsets are not continuations of each other.

\item

Eigenfunctions corresponding to the subsets $m_0^{(+R)}$ and $m_0^{(-R)}$ are the same, $\Psi^{0,R}_M(L;m_0^{(+R)})=\Psi^{0,R}_M(L;m_0^{(-R)})$.


\end{enumerate}

\item

The subset $\Psi^{1,R}_M(L;m^{(R)})$ of linearly independent regular  eigenfunctions corresponds to the spectrum of $m$ 
which is symmetric under the reflection of the sign: $m^{(R)}=m_1^{(R)}=m_1^{(-R)} \, {\rm U} \, m_1^{(+R)}$;

\begin{enumerate}
\item
Same as 2(a) above.


\item
The minimal value in the set $m_1^{(R)}$ is $m_{1min}^{(R)}=m_{1min}^{(-R)}=-L+1$ and the maximal value is $m_{1max}^{(R)}=m_{1max}^{(+R)}=L-1$.

\item

Same as 2 (c) above.

\item

Eigenfunctions corresponding to the subsets $m_1^{(+R)}$ and $m_1^{(-R)}$ are the same, $\Psi^{1,R}_M(L;m_1^{(+R)})=\Psi^{1,R}_M(L;m_1^{(-R)})$.


\end{enumerate}
\end{enumerate}

Condition of reducing hypergeometric functions to polynomials generates singular functions as well. Since our goal is to identify and describe the subset of regular functions, we do not give explicit details of the subset of singular functions, just remark that for the $L$ integer singular functions are not generated in the sequence $\{\Psi(L,-L),\,\Psi(L,-L+2),\cdots,\Psi(L,L-2),\,\Psi(L,L)\}$, and appear in the sequence $\{\Psi(L,-L),\,\Psi(L,-L+1),\cdots,\Psi(L,L-1),\,\Psi(L,L)\}$.

Lastly we consider the case $2\beta=|m|$. This paremeterisation leads to a different picture. The linearly independent solutions are now expressed as
\begin{eqnarray}
&& \Psi_M(L;2\beta)=\Psi^0_M(L;|m|)= (1-\xi^2)^{|m|/2} { }_2 F_1 \left( \frac{1}{2}+ \frac{|m|}{2}+ \frac{L}{2}, \frac{|m|}{2}- \frac{L}{2}; \frac{1}{2};\xi^2\right), \nonumber \\
&& \Psi_M(L;2\beta)=\Psi^1_M(L;|m|)= \xi (1-\xi^2)^{|m|/2} { }_2 F_1 \left( 1+\frac{|m|}{2}+ \frac{L}{2},\frac{1}{2}+\frac{|m|}{2}- \frac{L}{2}; \frac{3}{2};\xi^2\right);
\label{eq19}
\end{eqnarray}
In distinct of parameterisation $2\beta=\pm m$, now only two conditions for getting polynomials remain: 
 $|m|/2-L/2=-k$ for $\Psi^0_M(L;|m|)$ and $|m|+1/2-L/2=-k$ for $\Psi^1_M(L;|m|)$.
By applying these conditions, $|m|=L-2k\geq 0$ and $|m|=L-1-2k\geq 0$, only regular functions and the corresponding spectra are obtained. The singular functions and corresponding spectra are not generated since $|m|>0$. 
By enumerating integer values of $k$-parameter one enumerates all positive as well as all negative values of the spectrum of $m$. 
That is, moving with the steps size 2 one is not transferred from the negative values of the spectrum to the positive ones and vice versa but rather both parts are united into one quantity $|m|$, and both parts of the spectrum, with the opposite signs, are simultaneously enumerated. 
Therefore, to find regular functions as solutions in the eigenvalue/eigenfunction problem of the square of the angular momentum, the most convenient parameterisation is $2\beta=|m|$.


The sets of eigenvalues $m_0^{(R)}$ and $m_1^{(R)}$ and their corresponding eigenfunctions, corresponding to the parameterisation $2\beta=\pm m$, can be formally united in one set. Using numerical ordering from the smallest to the largest this united set is presented as a following sequence:
\begin{eqnarray}
m^{(R)} &=& \{ m_0^{(R)} {\rm U} \, m_1^{(R)} \} = \{-L;-L+1;-L+2;\cdots ; m_{max}^{(-R)} ; m_{min}^{(+R)};\cdots;L-2,L-1;L \} \nonumber\\
m_{max}^{(-R)} &=& -m_{min}^{(+R)},
\label{eq1.20}
\end{eqnarray}
where, depending on a numeric value of $L$, $m_{min}^{(+R)}$ is $(L-2[L/2])$ - the minimal positive value corresponding to $m_0^{(+R)}$, or $(L-1-2[(L-1)/2])$ - the minimal positive value corresponding to $m_1^{(+R)}$. 
The set of regular eigenfunctions corresponding to ordering (\ref{eq1.20}) is:
\begin{eqnarray}
\Psi_M(L;m) &=& \left\{ \Psi^{0, R}_M(L;-L); \Psi^{1, R}_M(L;-L+1); \Psi^{0,R}_M(L;-L+2);\cdots ; \Psi^R_M(L;m_{max}^{(-R)}); \right. \nonumber\\
&& \left. \Psi^R_M(L;m_{min}^{(+R)}); \cdots ;\Psi^{0,R}_M(L;L-2);\Psi^{1,R}_M(L;L-1); \Psi^{0, R}_M(L;L)\right\}.
\label{eq1.21}
\end{eqnarray}
Both sets (\ref{eq1.20}) and (\ref{eq1.21}) are obtained by merging two sequences with steps size 2 such that each of them becomes a sequence with the step size 1. Conditions of getting polynomials that lead to $m_0^{(R)}$ and $\Psi^0_M(L;m_0^{(R)})$, are not compatible with the conditions of getting polynomials that lead to 
$m_1^{(R)}$ and $\Psi^1_M(L;m_1^{(R)})$. Therefore in the above introduced sets with steps size 1 the functions $\Psi^0_M(L;m_0^{(R)})$ and $\Psi^1_M(L;m_1^{(R)})$ are not merged in one set by conditions of getting polynomials or by some similar 
condition corresponding to any common feature, but rather only by the formal requirement to present the merged set as one with the step size 1. That is the reason why above we used the term "formal" for the sets (\ref{eq1.20}) and (\ref{eq1.21}).

Note that the maximum and minimum values of $m$, $m_{max,min}=\pm L$, reside in the spectrum $m_0^{(R)}$, correspondingly the eigenfunction 
$\Psi_M(L,\pm L)$ is regular only when from the two linearly independent functions $\Psi_M^{0}$ and $\Psi_M^1$ the $\Psi_M^{0,R}(L,\pm L)$ is choosed as an eigenfunction. Second solution is singular, $\Psi_M(L,\pm L)=\Psi_M^{1,S}(L,\pm L)$. Thus the sequence of regular eigenfunctions (\ref{eq1.21}) starts from $\Psi_M^0(L,-L)$ and ends with $\Psi_M^0(L,L)$. Moving with the steps size 2 up from $\Psi_M^0(L,-L)$ or down from $\Psi_M^0(L,L)$ we obtain sequences of functions $\Psi_M^0(L,-L+2k)$ and $\Psi_M^0(L,L-2k)$. The functions from these sequences satisfy $\Psi_M^0(L,-L+2k)=\Psi_M^0(L,L-2k)$. 

It is important to note that in establishing the subset of regular eigenfunctions of the operator of the square of the orbital momentum no constraint arises on the angular momentum $L$. Indeed, reducing hypergeometric function to a polynomial is possible for any $L$, integer as well as non integer. As an example, the subset of regular eigenfunctions (\ref{eq1.21}) for the case $L=2$ is comprised from functions with $m\in \{ -2,-1,0,1,2\}$ and for $L=2.2$ from functions with $m\in\{-2.2,-1.2,-0.2,0.2,1.2,2.2\}$.

\section{Analysis of the spectrum of eigenvalues generated by the requirement for the eigenfunctions to be regular}
\label{secIII}

As shown in section \ref{secII}, equation (\ref{eq1.4}) 
 has two linearly independent 
solutions $\Psi^0_{M}(\xi|L;m) = \Psi^0_{M}(-\xi|L;m)$ and $\Psi^1_{M}(\xi|L;m) = -\Psi^1_{M}(-\xi| L,m) $. 
Clearly any linear combination of these functions 
\begin{eqnarray}
\Psi_{M}(\xi|L;m) = C_0 \Psi^0_{M}(\xi|L;m) + C_1 \Psi^1_{M}(\xi |L,m), 
\label{eq2.2}
\end{eqnarray}
where $C_0$ and $C_1$ are arbitrary numerical coefficients, is also a solution to the same linear differential equation (\ref{eq1.4}).

In addition to the purely mathematical attribute of linear differential equation, stating that solution can always be presented as a linear combination (\ref{eq2.2}), in quantum mechanics there exists an analogous 
physical condition, the principle of superposition. According to this principle if physical system can be in states described by regular wave functions
$\Psi_1^R$ and $\Psi_2^R$ then it can be also be in a state described by the wave function 
\begin{eqnarray}
\Psi^R = C_1^R \Psi_1^R + C_2^R \Psi^R_2.
\label{eq2.3}
\end{eqnarray}
Despite the similarity of Eqs. (\ref{eq2.2}) and (\ref{eq2.3}) there are substantial differences between these two relations. Namely,  Eq.~(\ref{eq2.3}) is a sum of a regular functions while there is no such a requirement for terms in Eq.~(\ref{eq2.2}) and indeed, as we have seen in previous section, depending on conditions on $L,\,m$, functions $\Psi^0_M,\,\Psi^1_M$ may be regular as well as singular. Also, in the physical principle of superposition $\Psi_1^R$ and $\Psi^R_2$ may correspond to the 
two different eigenvalues of the same observable, e.g. $\Psi_1^R=\Psi_m(\phi|m_1)=\exp(i m_1\phi)$ and $\Psi^R_2=\Psi_m(\phi|m_2)=\exp(i m_2\phi)$, while nothing similar is meant in Eq.~(\ref{eq2.2}). On the contrary, 
the necessary condition of mixing in Eq.~(\ref{eq2.2}) is that eigenvalues of the given quantity corresponding to both terms must be the same. Because of this restriction Eq.~(\ref{eq2.2}) is not a condition equivalent to Eq.~(\ref{eq2.3}) when the 
terms in Eq.~(\ref{eq2.2}) may be singular. Such a case is realised by the eigenfunctions of the square of the angular momentum $\Psi^0_M(\xi|L;m)$ and $\Psi^1_M(\xi|L;m)$. For the Eqs. (\ref{eq2.2}) and (\ref{eq2.3})
to be equivalent in the sense that Eq.~(\ref{eq2.3}) could be obtained from Eq.~(\ref{eq2.2}), these functions must be regular and from the analysis of the previous section we know that the functions 
$\Psi^0_M(\xi|L;m)$ and $\Psi^1_M(\xi|L;m)$, $|m|\in |m_0^R|=L-2k$ and $|m|\in |m_1^R|=L-1-2k$ cannot be regular at the same time. That is, $\Psi^0_M(\xi|L;m)$ is regular for the numerical value of $m\in m_0^R(L;k)$ and for the same value of $m$ 
$\Psi^1_M(\xi|L;m)$ is necessarily singular, i.e. un-normalisable, and vice versa, $\Psi^1_M(\xi|L;m)$ is regular for the numerical value of $m\in m_1^R(L;k)$ and for the same value of $m$ 
$\Psi^0_M(\xi|L;m)$ is necessarily singular, i.e. un-normalisable. Therefore, in presenting solution to the eigenvalue/eigenfunction problem for the angular momentum in the form (\ref{eq2.2}) some procedure must be employed in order to filter out the regular function from the combination (\ref{eq2.2}).

Let us recall that 
 the solution to the quantum mechanical problem of the angular momentum historically was presented in terms 
of the well known associated Legendre functions $P^\mu_\nu, \,Q^\mu_\nu$  (see, e.g. \cite{shiff}, \cite{messiah}).
These functions, being linear combinations of fundamental solutions $\Psi^0_M$ and $\Psi^1_M$, are not necessarily regular and for reducing them to a regular solution a procedure of filtering coefficients is used.

To clarify details of this filtering procedure let us consider expressions for the associated Legendre functions $P^\mu_\nu$ and $Q^\mu_\nu$  in terms of $\Psi^0_M$ and $\Psi^1_M$(see, e.g. Ref.~\cite{Abramowitz}): 
\begin{eqnarray}
&& \left[ P^\mu_\nu(\xi)/(-4)^{-|m|/2} \pi^{1/2}\right] |_{\nu=L; \mu=-|m|} = \left[ C_0 \Psi^0_{M}(\xi| L; |m|) + C_1 \Psi^1_{M}(\xi | L; |m| )\right],\nonumber\\
&& C_0 = \left[ \Gamma(1/2-L/2+|m|/2) \Gamma(1+L/2+|m|/2)\right]^{-1},
\nonumber\\
&& C_1 = -2 \left[ \Gamma(1/2+L/2+|m|/2) \Gamma(-L/2+|m|/2)\right]^{-1},\label{Gia}
\end{eqnarray}
and
\begin{eqnarray}
&& \left[ e^{i\mu\pi} Q^\mu_\nu (\xi)/(-4)^{-|m|/2} \pi^{1/2}\right] |_{\nu=L; \mu=-|m|} = \left[ C_3 \Psi^0_{M}(\xi | L; |m|) + C_4 \Psi^1_{M}(\xi | L; |m| )\right], \nonumber\\
&& 2 C_3 e^{\pm i(|m|+L+1)\pi/2} = \Gamma(1/2+L/2-|m|/2)/2 \Gamma(1+L/2+|m|/2), \nonumber\\
&& C_4e^{\pm i(|m|+L)\pi/2} = \Gamma(1+L/2-|m|/2)/\Gamma(1/2+L/2+|m|/2),
\label{eq2.4}
\end{eqnarray}
where $\Gamma(z)$ is the Euler gamma function \cite{Abramowitz}.
 Since $\Psi^0_M$ and $\Psi^1_M$ are not simultaneously regular for a fixed values of $L$ and $|m|$, to obtain a regular solution we can use the following strategy: for the regular $\Psi^0_M$ coefficient in front of $\Psi^1_M$ must vanish and vice versa. This is the filtering procedure mentioned above.

As shown in the previous section, conditions for solutions to be regular result in the following relations:
\begin{equation}
\label{mcond}
|m|=L-2k, \quad |m|=L-1-2k.
\end{equation}
Correspondingly, the mixing coefficients of Eqs.~(\ref{Gia}) and (\ref{eq2.4}) take the form:
\begin{eqnarray}
&& C_0(L;|m|)|_{|m|=L-2k} = \left[ \Gamma(1/2-L+k) \Gamma(1+k)\right]^{-1},\nonumber\\
&& C_1(L;|m|)|_{|m|=L-2k} = -2 \left[ \Gamma(1/2+k) \Gamma(-L+k)\right]^{-1},\nonumber\\
&& C_3(L;|m|)|_{|m|=L-2k} \sim \Gamma(1/2+k)/ \Gamma(1+L-k),\nonumber\\
&& C_4(L;|m|)|_{|m|=L-2k} \sim \Gamma(1+k)/ \Gamma(1/2+L-k),\nonumber\\
&& C_0(L;|m|)|_{|m|=L-1-2k} = \left[\Gamma(1-L+k) \Gamma(3/2+k)\right]^{-1},\nonumber\\
&& C_1(L;|m|)|_{|m|=L-1-2k} = -2 \left[ \Gamma(1+k) \Gamma(1/2-L+k)\right]^{-1},\nonumber\\
&& C_3(L;|m|)|_{|m|=L-1-2k} \sim \Gamma(1+k)/ \Gamma(3/2+L-k),\nonumber\\
&& C_4(L;|m|)|_{|m|=L-1-2k} \sim \Gamma(3/2+k) / \Gamma(1+L-k).
\label{eq2.5}
\end{eqnarray}
Associated Legendre functions are regular when the following filtering requirements are satisfied:
\begin{eqnarray}
&& C_1(L;|m|)|_{|m|=L-2k} = 0; \label{eq2.6.1a}\\
&& C_4(L;|m|)|_{|m|=L-2k} = 0; \label{eq2.6.1b}\\
&& C_0(L;|m|)|_{|m|=L-1-2k} =0; \label{eq2.6.2a}\\
&& C_3(L;|m|)|_{|m|=L-1-2k} =0; 
\label{eq2.6.2b}
\end{eqnarray}
It is seen from the explicit form of the mixing coefficients that the condition of Eq.~(\ref{eq2.6.1b}) cannot be satisfied. Condition (\ref{eq2.6.1a}) is satisfied if $L$ is a non-negative integer number. 
Similarly, condition (\ref{eq2.6.2b}) cannot be satisfied, while the condition (\ref{eq2.6.2a}) is satisfied for a non-negative integer $L$. Associated Legendre functions are regular, i.e. admissible only when $L$ is a non-negative integer; otherwise they are singular, i.e. non admissible.

Let us demonstrate this on a concrete numerical examples. 
We consider following values of $L,\,m$: $L=\{ 2;3/2;1.2\}$ and $m=\{\pm2; \pm3/2; \pm1.2\}$. For the functions and mixing coefficients of Eq.~(\ref{eq2.5}) we obtain:
\begin{eqnarray}
&& \Psi_M^0(L=2;|m|=|\pm 2|)=\Psi_M^0(2;2)^R,\ \Psi_M^1(L=2;|m|=|\pm 2|)=\Psi_M^1(2;2)^S; \nonumber\\
&& C_0(2;|\pm 2) = \left[ \Gamma\left(-3/2\right) \Gamma(1)\right]^{-1}\neq 0; \nonumber\\
&& C_1(2;|\pm 2) = -2 \left[ \Gamma\left(1/2\right) \Gamma(-2)\right]^{-1}=0; \ P^2_2(\xi)=P^2_2(\xi)^R \nonumber\\
&& C_3(2;|\pm 2) \sim \Gamma\left(1/2\right) /\Gamma(3)\neq 0; \nonumber\\
&& C_4(2;|\pm 2) \sim \Gamma\left(3/2\right)/\Gamma(3)\neq 0; \ \ \ Q^2_2(\xi)=Q^2_2(\xi)^S \nonumber\\
&& \nonumber\\
&& \Psi_M^0(L=3/2;|m|=|\pm 3/2|)=\Psi_M^0(3/2;3/2)^R; \ \Psi_M^1(L=3/2;|m|=|\pm 3/2|)=\Psi_M^1(3/2; 3/2)^S; \nonumber\\
&& C_0(3/2;|\pm 3/2) = \left[ \Gamma\left(-1\right) \Gamma(1)\right]^{-1}= 0; \nonumber\\
&& C_1(3/2;|\pm 3/2) = -2 \left[ \Gamma\left(1/2\right) \Gamma(-3/2)\right]^{-1}\neq 0; \ P^{3/2}_{3/2}(\xi)=P^{3/2}_{3/2}(\xi)^S \nonumber\\
&& C_3(3/2;|\pm 3/2) \sim \Gamma\left(1/2\right) /\Gamma(5/2)\neq 0; \nonumber\\
&& C_4(3/2;|\pm 3/2) \sim \Gamma\left(1\right)/\Gamma(2)\neq 0; \ \ \ Q^{3/2}_{3/2}(\xi)=Q^{3/2}_{3/2}(\xi)^S \nonumber\\
&& \nonumber\\
&& \Psi_M^0(L=1.2;|m|=|\pm 1.2|)=\Psi_M^0(1.2;1.2)^R; \ \Psi_M^1(L=1.2;|m|=|\pm 1.2|)=\Psi_M^1(1.2;1.2)^S; \nonumber\\
&& C_0(1.2;|\pm 1.2) = \left[ \Gamma\left(0.7\right) \Gamma(1)\right]^{-1}\neq 0; \nonumber\\
&& C_1(1.2;|\pm 1.2) = -2 \left[ \Gamma\left(1/2\right) \Gamma(-1.2)\right]^{-1}\neq0; \ P^{1.2}_{1.2}(\xi)=P^{1.2}_{1.2}(\xi)^S \nonumber\\
&& C_3(1.2;|\pm 1.2) \sim \Gamma\left(1/2\right) /\Gamma(2.2)\neq 0; \nonumber\\
&& C_4(1.2;|\pm 1.2) \sim \Gamma\left(1\right) /\Gamma(1.7)\neq0; \ \ \ Q^{1.2}_{1.2}(\xi)=Q^{1.2}_{1.2}(\xi)^S.
\label{eqnonumber}
\end{eqnarray}
Analogous results are obtained for the other values of $m$. For a non-negative integer $L$ the functions $P^{-|m|}_L(\xi)$ are regular and functions $Q^{-|m|}_L(\xi)$ remain singular. For an integer values of $L$ the set of Eq.~(\ref{eq1.20}) becomes 
a well known set (see, e.g. Ref.~\cite{shiff}):
\begin{eqnarray} 
&& m_{max}^{(-R)} =-m_{min}^{(+R)} =0; \ m^{(R)}= \{ -L; -L+1;-L+2; \cdots ; 0; \cdots; L-2;L-1;L \}; \nonumber\\
&& L=\{ 0; 1; 2; \cdots; \infty \};
\label{eq2.7} 
\end{eqnarray}

Thus, if one chooses to present solution of the eigenvalue problem of the angular momentum in terms of associated Legendre functions then from the requirement that eigenfunction has to be regular it follows that $L$ is necessarily integer and the spectrum of $m$ is given by Eq.~(\ref{eq2.7}). 
In doing so some set of the regular functions and corresponding non-integer values of $L$ disappear from the solution. 

But there is no any mathematical or physical argument or requirement that would dictate that the solution of the eigenvalue equation Eq.~(\ref{eq1}) should necessarily be presented in form (\ref{eq2.2}). 

As it is shown in the previous section, $\Psi_M^0,\,\Psi_M^1$ are the linearly independent solutions of Eq.~(\ref{eq1}) and, depending on which regularity condition out of Eq.~(\ref{mcond}) is realized, one of these two functions 
becomes a regular function and can be chosen as a solution to the eigenvalue problem of the square of the orbital momentum. This choice, from the theoretical point of view, is no worse and no better than choice in terms of $P^\mu_\nu,\,Q^\mu_\nu$. Solution presented by only $\Psi_M^0$ or by only $\Psi_M^1$ is regular and in distinct of presenting solution in terms of $P^\mu_\nu,\,Q^\mu_\nu$, does not generate any constraint on $L$; regular, i.e. physically admissible solutions exist for a non-integer $L$ as well.

Therefore, we conclude that the spectrum of Eq.~(\ref{eq2.7}) is just an artefact of 
presenting eigenfunctions in the form of Eq.~(\ref{eq2.2}).


Different approach used to demonstrate that the spectrum is given by  Eq.~(\ref{eq2.7}), i.e. that $L$ acquires only integer values, is based on the analysis of commutation relations of the angular momentum operators. 
In the next section we address these arguments. 


\section{Analysis of the spectrum of eigenvalues generated by the commutation relations of the angular momentum operators}
\label{secIV}

Let as analyze the reasoning based on the commutation relations used to demonstrate that the eigenvalues of the 
square of the angular momentum and its third component can be only integer numbers (see, e.g. \cite{shiff}, \cite{messiah}). It is formulated as follows:
if $|L,m>$ is a normalisable state vector satisfying 
\begin{equation}
\hat M^2|L;m> = L(L+1) |L;m>; \ \ \hat M_z|L;m> = m |L;m>,
\label{eq3.1}
\end{equation}
then the following mathematical relations must hold (see e.g. Ref.~\cite{shiff}, section XIII):

\begin{enumerate}
\item
$-L\leq m\leq L$;

\item
If $m=L$ then $\hat M^+|L;L>=0$, where $\hat M^+=\hat M_x+i \hat M_y$, \\if $m=-L$ then $\hat M^-|L;L>=0$, where $\hat M^-=\hat M_x-i \hat M_y$.

\item
If $m\neq L$ then $\hat M^+|L;m>$ is an eigenvector with eigenvalues of the angular momentum $L(L+1)$ and $(m+1)$,\\if $m\neq -L$ then $\hat M^-|L;m>$ is an eigenvector with eigenvalues of the angular momentum $L(L+1)$ and $(m-1)$.



\item
If $ (\hat M^{\pm})^p|L;m>\neq 0$ then it is an eigenvector with eigenvalues of the square of angular momentum and its third momentum correspondingly $L(L+1)$ and $(m\pm p)$.

\item
In the sequences of eigenvectors $ \hat M^+ |L;m>; (\hat M^{+})^2|L;m>; \cdots ; (\hat M^{+})^p|L;m>$ and $ \hat M^- |L;m>; (\hat M^{-})^2|L;m>; \cdots ; (\hat M^{-})^q|L;m>$ there always can be found such values of $p$ and $q$ 
that the following two relations simultaneously hold:
\begin{equation}
\label{mpq}
m+p=L,\quad m-q=-L,
\end{equation}
i.e. acting repeatedly on any $|L;m\rangle$ with $\hat M^+$ and $\hat M^-$ we obtain both $|L;L\rangle$ and $|L;-L\rangle$. Consequently, as $p$ and $q$ are positive integer numbers, the difference $(m+p)-(m-q)=(p+q)=2L$ is also an integer number. 

\end{enumerate}
After that it is assumed that the quantization of the angular momentum is proven.

To analyze proof of quantization based on the statements 1-5, let us first note 
that from the theoretical standpoint if any argument is not determined either by commutation relations or by physical requirements then 
the condition stated in this argument is not necessarily to hold. 
Such an argument, relevant for our discussion is the point 5 above. We demonstrate that point 5 is not always valid 
on the example with the three numerical values of $L$: $L=2$ - corresponding to integer values, $L=3/2$ - corresponding to half-integer values, and $L=1.2$ - non-integer value, which is also not a half-integer. Let us assume at this stage that the conditions of items 1-5 are indeed satisfied and consider the following eigenvalues of $L$: $\{ 2;3/2;1.2 \}$. Since Eq.~(\ref{eq1.4}) contains $m$ quadratically, both $\Psi_{Mm}(L;m)$ and $\Psi_{Mm}(L;-m)$ are solutions to this equation. Eigenfunctions $\Psi_{Mm}(2;\pm 2)$, $\Psi_{Mm}(3/2; \pm 3/2)$ and $\Psi_{Mm}(1.2;\pm 1.2)$ are regular, i.e. normalizable solutions of Eq.~(\ref{eq1.4}). 
Then, according to the points 1-5 above, the following sequences will also be eigenfunctions:
\begin{eqnarray}
&& \{ \hat M^{\pm}\Psi_{Mm}(2;\mp 2),\,(\hat M^{\pm})^2\Psi_{Mm}(2;\mp 2),\,(\hat M^{\pm})^3\Psi_{Mm}(2;\mp 2),\,(\hat M^{\pm})^4\Psi_{Mm}(2;\mp2)\},\nonumber\\
&& \{ \hat M^{\pm}\Psi_{Mm}(3/2;\mp 3/2),\,(\hat M^{\pm})^2\Psi_{Mm}(3/2;\mp 3/2),\,(\hat M^{\pm})^3\Psi_{Mm}(3/2;\mp 3/2) \}, \nonumber\\
&& \{ \hat M^{\pm}\Psi_{Mm}(1.2;\mp 1.2),\,(\hat M^{\pm})^2\Psi_{Mm}(1.2;\mp 1.2);\cdots \}. \nonumber
\end{eqnarray}
The $(L;m)$ values corresponding to these sequences of functions are:
\begin{eqnarray}
&& L=2; \ \ \ \ \ m\downarrow=\{ 2;1;0;-1;-2\}; \ \ \ \ \ \ \ \ \ \ \ m\uparrow=\{-2;-1;0;1;2\}; \nonumber\\
&&L=3/2; \ \ m\downarrow=\{ 3/2;1/2;-1/2;-3/2\}; \ \ m\uparrow=\{-3/2;-1/2;1/2;3/2\}; \nonumber\\
&& L=1.2; \ \ \ m\downarrow=\{ 1.2;0.2;-0.8;\cdots \}; \ \ \ \ \ \ \ m\uparrow=\{-1.2;-0.2;0.8;\cdots \}.\nonumber\\
\end{eqnarray}

As seen from these expressions, in case $L=2$ by acting on functions $\Psi_{Mm}(L;\pm L)$ with the operators $\hat M^+$ and $\hat M^-$, two identical sets are generated: $m\downarrow$, obtained by applying $\hat M^-$ to $\Psi_{Mm}(2;2)$, and $m\uparrow$, obtained by applying $\hat M^+$ to $\Psi_{Mm}(2;-2)$.
These sets have a property that moving down with the unit step from the value $m_{max}=2$ we arrive to the minimal value $m_{min}=-2$ and vice versa, i.e. from the value $m_{min}=-2$ we move up to the maximal 
value $m_{max}=2$. For $L=3/2$ we get the same result - sets $m\uparrow$ and $m\downarrow$ are identical.


In the case of $L=1.2$ the sets $m\downarrow=\{ 1.2;0.2;-0.8; \cdots \}$, generated by $\hat M^-$, and $m\uparrow=\{ -1.2;-0.2; 0.8; \cdots \}$, generated by $\hat M^+$, differ and have no intersection. Starting with any element of $m\downarrow$ by repeatedly acting with $\hat M^+$ we  arrive at $m_{max}=1.2$, however by repeatedly acting with $\hat M^-$ we do not arrive at the minimal value in $m\uparrow$, to 
$m_{min}=-1.2$. Hence, in case of a non-integer $L$ point 5, stating that for the sequences of functions generated by acting with $\hat M^-$ and $\hat M^+$ one necessarily finds such integers $p$ and $q$ that 
starting with some some $\Psi_{Mm}(L;m)$ of this sequence, by acting with  $(\hat M^+)^p$ and $(\hat M^-)^q$ one simultaneously obtains $\Psi_{Mm}(L;L)$ as well as $\Psi_{Mm}(L;-L)$, is no longer valid.
In case of the non-integer values of $L$, if by acting with $(\hat M^+)^p$ on $\Psi_{Mm}(L;m)$ we obtain $\Psi_{Mm}(L;L)$, then this state necessarily belongs to the spectrum of type $m\uparrow$ and by acting on it with $(\hat M^-)^q$ we cannot obtain
$\Psi_{Mm}(L;-L)$. Similarly, if by acting with $(\hat M^-)^q$ on $\Psi_{Mm}(L;m)$ we obtain $\Psi_{Mm}(L;-L)$ then this state necessarily belongs to the spectrum of type $m\downarrow$ and by acting on it with $(\hat M^+)^p$ we cannot obtain
$\Psi_{Mm}(L;L)$. In other words Eq.~(\ref{mpq}) is no longer valid and consequently, angular momentum quantization can not be proved.

Let us ask where the requirements formulated in point 5 above come from, what are they based upon.
These requirement, namely that by repeatedly applying operator $\hat M^{+}$ to $\Psi_{Mm}(L;m)$ we arrive to $\Psi_{Mm}(L;L)$, and, repeatedly applying operator $\hat M^-$ to the same $\Psi_{Mm}(L;m)$ we arrive to $\Psi_{Mm}(L;-L)$, do not follow neither from commutation relations nor from any physical arguments.

Consequently we conclude that in the framework of the algebra of commutation relations the conditions stated in points 5 above is not the one which must be necessarily satisfied Violation of the requirement in point 5 does not contradict to any physical requirement or commutation relations.

Therefore the integer, half-integer and also any other real values of $L$ are compatible with the algebra of commutation relations as well as with physical requirements.

We finish this section by listing how, by acting with operators $\hat M^{\pm}$ upon $\Psi_{Mm}^0,\,\Psi_{Mm}^1$, results move from the set of regular functions to set of singular functions, vice versa, or remain in the original set. First let us recapitulate result from section \ref{secII} stating that if  $\Psi^0_{Mm}(L;m)$ is a regular function for fixed $L$, then $\Psi^1_{Mm}(L;m)$ is necessarily singular and vice versa.

We omit lengthy straightforward calculations and just state the results:
\begin{enumerate}
\item
For any $L$, if $m$ and $m\pm 1$ belong to the spectrum with the same sign, that is if both $m$ and $m\pm 1$ are positive or both $m$ and $m\pm 1$ are negative, then $\hat M^{\pm}\Psi_{Mm}(L;m)^R\sim \Psi_{Mm}(L;m\pm 1)^R$ and  $\hat M^{\pm}\Psi_{Mm}(L;m)^S\sim \Psi_{Mm}(L;m\pm 1)^S$. In other words, in this case if $\Psi$ is regular (singular), the  $\hat M^{\pm}\Psi$ is also regular (singular).

\item
If $L$ is non-integer, $L\neq[L]$ and $m$ and $(m\pm 1)$ belong to spectra with different signs, then $\hat M^{\pm}\Psi_{Mm}(L;m)^R\sim\Psi_{Mm}(L;m\pm 1)^S$. In other words, operators $\hat M^{(\pm)}$ bring the regular functions $\Psi_{Mm}(L;m)^R$ to singular functions $\Psi_{Mm}(L;m\pm 1)^S$.
\end{enumerate}
Along the statements above an important feature is that 
$m|_{(-\infty)}^{(L)}$ and $m|^{(+\infty)}_{(-L)}$ numerical sequences have no intersection unless $L$ is integer or half-integer. The case of integer $L$ is distinguished by the fact that instead of statement 2 we now have:

3. If $L$ is integer, $L=[L]$ and $m$ and $(m\pm 1)$ belong to spectra with different signs, then $\hat M^{\pm}\Psi_{Mm}(L;m)^R\sim \Psi_{Mm}(L;m\pm 1)^R$. In other words, operators $\hat M^{(\pm)}$ bring regular functions $\Psi_{Mm}(L;m)^R$ to regular functions $\Psi_{Mm}(L;m\pm 1)^R$.

The case of half-integer $L$ is described by statement 2, corresponding to non-integer numbers:
\begin{eqnarray}
&& \hat M^- \Psi^0_{Mm}(L;1/2)^R|_{[L]=2k} = (L+1/2)^2 \Psi^1_{Mm}(L;-1/2)^S|_{[L]=2k}; \nonumber \\ 
&& \hat M^+ \Psi^1_{Mm}(L;-1/2)^S|_{[L]=2k} = - \Psi^0_{Mm}(L;1/2)^R|_{[L]=2k}; \nonumber \\ 
&& \hat M^- \Psi^1_{Mm}(L;1/2)^R|_{[L]=2k+1} = \Psi^0_{Mm}(L;-1/2)^S|_{[L]=2k+1}; \nonumber \\ 
&& \hat M^+ \Psi^0_{Mm}(L;-1/2)^S|_{[L]=2k+1} = (L+1/2)^2 \Psi^1_{Mm}(L;1/2)^R|_{[L]=2k+1}; \nonumber \\ 
&& \hat M^+ \Psi^0_{Mm}(L;-1/2)^R|_{[L]=2k} = (L+1/2)^2 \Psi^1_{Mm}(L;1/2)^S|_{[L]=2k}; \nonumber \\ 
&& \hat M^- \Psi^1_{Mm}(L;1/2)^S|_{[L]=2k} = - \Psi^0_{Mm}(L;-1/2)^R|_{[L]=2k}; \nonumber \\ 
&& \hat M^+ \Psi^1_{Mm}(L;-1/2)^R|_{[L]=2k+1} = \Psi^0_{Mm}(L;1/2)^S|_{[L]=2k+1}; \nonumber \\ 
&& \hat M^- \Psi^0_{Mm}(L;1/2)^S|_{[L]=2k+1} = (L+1/2)^2 \Psi^1_{Mm}(L;-1/2)^R|_{[L]=2k+1}; 
\label{eq3.10}
\end{eqnarray}
Correspondingly, for the non-integer $L$ by acting with the operators $\hat M^{(\pm)}$ the set of singular functions is attached to the set of regular functions (for the half-integer $L$ this fact is also mentioned in Refs.~\cite{Pauli}, \cite{mer})
and this attachment is transparent in both directions, that is, by acting with operators $\hat M^{(\pm)}$ we move from singular functions to regular ones and vice versa. 

In connection with the above results following important comment is in order:
as shown in section \ref{secII}, obtaining regular eigenfunctions and the spectrum via condition of reducing hypergeometric expressions for $\Psi^
{0,1}_{Mm}$ to polynomials, singular eigenfunctions do not appear at 
all when the parameterization $2\beta\equiv\sqrt{m^2}=|m|$ is used.    However, when using operators $\hat M^{(\pm)}$ to establish the spectrum as it is done, e.g. in \cite{wein}, singular functions are generated even for the parameterization $2\beta=|m|$, since the operators $\hat M^{\pm}$, containing $d/d\xi$, lower the exponent $\beta$ in the expression $\Psi=(1-\xi^2)^{\beta}\,F$ and acting repeatedly with $\hat M^{\pm}$ results in a singular function of $\xi$. If for establishing the set of normalizable eigenfunctions one uses procedure described in section \ref{secII} and not the method based on using operators $\hat M^{\pm}$, the singular functions do not appear in the set of eigenfunctions.

In the quantum-mechanical problem of the angular momentum operators $\hat M^{\pm}$ are often  treated on the same footing as operators corresponding to observable and the 
results of mathematical operations connected with their actions are considered as the conditions which have to be satisfied. For example, in Ref.~\cite{wein} the set of normalizable eigenfunctions is defined 
as the subset of functions obtained from $\Psi_L(L;\pm L)$ by acting with $\hat M^{(\pm)}$. The reason for giving to these operators such a high status stems from the paper by Pauli \cite{Pauli} in which the the issue of non-uniqueness of the eigenfunctions of the angular momentum operator is addressed. There is however no reason for giving these operators such a special status as they do not belong to a complete set of commuting operators. 

The analysis of the quantization of eigenvalues is connected not only to the properties of eigenfunctions of the square of the angular momentum but also to the non-uniqueness of the eigenfunctions of the third component of the angular momentum. Therefore arguments by Pauli will be 
addressed in our next publication where we will consider the issue wether $m$ is only integer by analyzing the properties of the eigenfunctions and eigenvalues of the third component of the angular momentum. 

\section{Conclusions}
\label{secVI}

From the above analysis we conclude:

\begin{itemize}
\item

A set of eigenfunctions of the operator of the square of orbital momentum, $\Psi_{Mm}(\xi,\phi |L;m)$, consists of subsets of singular and non-singular functions.

\item

Eigenvalues of the square of the angular momentum corresponding to non-singular, i.e. normalizable eigenfunctions can be integer as well as non-integer. 

\item

The main statement of the analysis of the solutions to the eigenvalue/eigenstate equations, citied in textbooks of quantum mechanics, that the only integer eigenvalues are admissible, is an artefact of 
considering a specific linear combination of linearly independent solutions (realised as an associated Legendre functions, the so called Spherical Harmonics). This requirement is neither a physical nor a mathematical necessary condition.


\item
If the condition of normalisability of eigenfunctions is realised by imposing conditions of getting polynomials then the subset of singular functions does not appear when the parameterization $(m^2)^{1/2}=|m|$ is chosen. This parameterization preserves in the expressions of eigenfunctions the symmetry present in the initial equation 
$\hat M^2(m)=\hat M^2(-m)$.



\item
In the operator formalism the set of physical states is completely factorised from the set of singular functions only in the case when $L$ is integer.

\item 
In the operator formalism the statement that $L$ can be only integer is a result of assigning to the operators $\hat M^{(\pm)}$ higher status than it follows from the principles of quantum mechanics. 

\item
To guarantee that $L$ acquires only integer values one either needs to find additional arguments on top to those which are usually stated when using the mechanism of solving the eigenvalue problem 
of the angular momentum and/or applying the algebra of commutation relations, or alternatively, one has to admit that in the framework of quantum mechanics $L$ may be integer as well as non-integer.

\end{itemize}

We are indebted with J.~T.~Gegelia for useful discussions and critically reading the manuscript.


\end{document}

\bibitem{Merzbacher} 
E.~Merzbacher, "Single Valuedness of Wave function". American Journal of Physics, Vol.30, N4 (1962).

\bibitem{Blokhintsev} 
D.~I.~Blokhintsev, "Principles of quantum mechanics", Allyn and Bacon, Boston, MA, 1964.

\bibitem{Landau:1991wop} 
L.~D.~Landau and E.~M.~Lifshits,
``Quantum Mechanics: Non-Abramowitz}
\bibitem{Abramowitz}
M.~Abramowitz and I.~Stegun, "Handbook of Mathematical Functions", Dover, New York, 1970. Abramowitz}
\bibitem{Abramowitz}
M.~Abramowitz and I.~Stegun, "Handbook of Mathematical Functions", Dover, New York, 1970.,''
Butterworth-Heinemann, Oxford", 1991. 

\bibitem{Fock} 
V.~A.~Fock, "Fundamentals of Quantum Mechanics", Mir Publishers; 1978. 